\documentclass[aps,prc,twocolumn,showpacs]{revtex4}
\usepackage{graphicx,epsfig}
\newcommand{ \be }{\begin{equation}}
\newcommand{ \ee }{\end{equation}}
\newcommand{ \bea }{\begin{eqnarray}}
\newcommand{ \eea }{\end{eqnarray}}
\newcommand{ \la }{\langle}
\newcommand{ \ra }{\rangle}

\newcommand{ \CP }{$\cal{C}\cal{P}$}
\newcommand{ \cP }{$\cal{P}$}

\usepackage{color}

\begin{document}

\title{
Parity violation in hot QCD: how to detect it
}

\author{Sergei A.~Voloshin}

\affiliation{Department of Physics and Astronomy, 
Wayne State University, Detroit, Michigan 48201}

\date{\today} 

\begin{abstract}
In a recent paper (arXive:hep-ph/0406125) entitled {\em Parity violation in hot
QCD: why it can happen, and how to look for it}, D.~Kharzeev argues for
the possibility of \cP- and/or \CP- violation effects in heavy-ion
collisions, the effects that can manifest themselves via
asymmetry in $\pi^{\pm}$ production with respect to the direction of the 
system angular momentum.
Here we present an experimental observable that can be used to
detect and measure the effects. 
\end{abstract}

\pacs{11.30.Qc, 12.38.Qk, 25.75.Ld, 25.75.Nq} 

\maketitle

%\section{Introduction}
The possibility of strong \cP- and \CP-violation in heavy ion collisions
has been proposed first in~\cite{kharzeev1}. 
Different experimental observables sensitive to the presence
of \cP- and/or \CP-odd domains in the deconfined QCD vacuum
have been already discussed in the original papers and later 
in~\cite{voloshin,starP}. 
Remarkably, all the observables which have been discussed are related
in smaller or larger extent
to the anisotropic flow study efforts. 
In general, \cP- and \CP-symmetry violation effects proposed 
in~\cite{kharzeev1} manifest itself via a non-statistical difference 
of the reaction planes reconstructed
using different groups of particles, either of different charge, or  
in different kinematic regions.
In symmetric nuclear collision (only those are discussed in this note)
there should be only one plane of symmetry, and therefore any 
observation of the opposite 
would mean \cP- and/or \CP- violation effects.        
Interestingly, many of the 'symmetry sensitive' quantities are routinely 
calculated in flow analyses for  'quality assurance' purposes (checking
analysis consistency). No deviation from expectations based on 
symmetry with respect to the reaction plane has been observed so
far.  

However, refs.~\cite{kharzeev1,voloshin,starP} do not discuss one important 
case, namely
the possibility of preferential emission of particle/antiparticle,
e.g. $\pi^\pm$, into opposite sides of the reaction plane. 
This happens to be exactly the observable signal
of the \cP- and \CP-breaking mechanism
discussed by Kharzeev in his recent preprint~\cite{kharzeev2}.
Kharzeev argues that due to the parity violating interactions, the
asymmetry in pion production along the direction of the system angular
momentum (perpendicular to the reaction plane)  could be as high as 
of the order of 
one percent in midcentral Au+Au collisions at RHIC. 
The orientation of the asymmetry (parallel or anti-parallel to the
direction of the angular momentum) can change from event 
to event, and therefore the effect can be detected only by correlation study.

In this short note we propose to use for that purpose a technique 
that is well known in
anisotropic flow analysis and usually referred to as  mixed 
harmonics technique~\cite{method} 
or three particle correlations~\cite{olli}.
The essence of this technique is just in the isolation of 
correlations related  to a given direction. 
Suppose that positive pions are emitted
preferentially in positive $y$ direction (along the angular momentum).
The azimuthal distribution in this case can be written as
 $dN/d\phi \propto (1+2a \sin(\phi))$, where
$\phi$ is the particle emission azimuthal angle relative to the
reaction plane ($\Psi_{RP}$), and the
parameter $a$ can be directly related to the
asymmetry in pion production discussed in~\cite{kharzeev2}: 
$A_{\pi^+}=\pi a/4 \approx Q/N_{\pi^+}$. 
In the latter expression 
$Q$ is the topological charge ($Q\ge 1$) and $N_{\pi^+}$ is the
pion multiplicity in about one unit of rapidity~\cite{kharzeev2}. 
For midcentral
Au+Au collisions at RHIC $N_{\pi^+}\sim 100$ and 
these estimates yield a low limit on $a$ of the order of one percent.
Let us consider azimuthal correlation between particles $a$ and
$b$ by evaluating the quantity
\bea
&&
\la 
\cos(\phi_a-\Psi_2) \cos(\phi_b-\Psi_2) 
\nonumber \\
&& \hspace{25mm} -\sin(\phi_a-\Psi_2) \sin(\phi_b-\Psi_2) 
\ra
\label{emh}
\\
&&
=
\la \cos(\phi_a +\phi_b -2\Psi_2) \ra =
(v_{1,a}v_{1,b} - a_a a_b) \, \la \cos(2\Psi_2) \ra
\nonumber
\eea
where the average is taken over events, $\Psi_2$ is the second
harmonic event plane, $\la \cos(\Psi_2-\Psi_{RP}) \ra$ is 
the so called event plane
resolution (how well on average one reconstructs the reaction plane from
elliptic flow; for details see~\cite{method}).
The final expression reflects the correlations along the two axes, one in the
reaction plane (directed flow, characterized by 
$\la \cos(\phi-\Psi_{RP})\ra \equiv v_1$) and perpendicular to the
reaction plane -- the manifestation of symmetry breaking discussed
in~\cite{kharzeev2}. All other correlations, being not sensitive to
the orientation of the reaction plane, cancel out (for the systematic
uncertainty in this statement see~\cite{olli,starv1v4} and discussion below).
The proportionality to the reaction plane resolution reflects 
a decrease in correlations due to finite ability to resolve the true
reaction plane orientation.
If only one particle is used to determine the event plane the equation
reduces to
\be
\la \cos(\phi_a + \phi_b -2\phi_c) \ra 
= (v_{1,a}v_{1,b}- a_a a_b) \, v_{2,c},
\label{e3p}
\ee  
where the typical values of the parameter $v_{2,c}$, elliptic 
flow of particle of type $c$, is of the order of 0.04--0.05 for midcentral
collisions.
Equations~(\ref{emh}) and (\ref{e3p}) are usually employed for
directed flow study~\cite{method,olli,starv1v4}. 
The main advantage of
these observables is their sensitivity to correlations in particle 
production along a given direction.
As already discussed above, these observables represent 
the difference in correlations
along the $x$ and $y$ axes, therefore any correlations that do not
depend on the orientation with respect to the reaction plane cancel
out. 
If directed flow is zero, the above observables present
a direct measure of the symmetry violation effects.
In relativistic heavy ion collision, the condition of $v_1=0$
can be achieved by studying the correlations in the rapidity region 
symmetric with respect to the mid-rapidity, such that the average
directed flow equals zero. 
As discussed in detail, for example, in~\cite{olli}, using this
technique
one is able to measure the correlations, $v_1$ or asymmetry 
parameter $a_{\pi^\pm}$ in our case,  with an accuracy at a sub-percent level.

Note the possibility of measuring the correlations using different
charge combinations: $\pi^+\pi^+$ and $\pi^-\pi^-$ correlations
should be negative, while $\pi^+\pi^-$ to be positive, and all
three to be of the same magnitude. These relations provide
an additional cross-check of the results if observed.

The main systematic uncertainty in three particle correlation
measurements is due to processes when particles $a$ and $b$ are
products of a resonance decay, and the resonance itself exhibits
elliptic flow~\cite{olli,starv1v4}.
Keeping only this contribution one can write:
\bea
&&
\la \cos(\phi_a + \phi_b -2\phi_c) \ra 
\nonumber
\\
&& \;\;\;\; =
\la \cos((\phi_a + \phi_b -2\phi_{res}) +2(\phi_{res}-\phi_c)) \ra
\nonumber
\\
&& 
\approx \frac{f_{res} \, \la  \cos(\phi_a + \phi_b -2\phi_{res}) \ra
\; v_{2,res}}{N_{\pi}} \, v_{2,c},
\eea
where $f_{res}$ is the fraction of pion pairs originating from resonance
decays (should be relatively small for the same charge combinations),
$ \la  \cos(\phi_a + \phi_b -2\phi_{res}) \ra $ can be considered as a
measure of the azimuthal
correlations of decay products with respect to the resonance azimuth,
$v_{2,res}$ is the resonance elliptic flow. The factor $1/N_\pi$ 
reflects the probability  that both pions in the pair are
from the same resonance.  Considering an estimate of such contribution
note that  $ \la  \cos(\phi_a + \phi_b -2\phi_{res}) \ra $ is zero if
the resonance is at rest, and become non-zero only due to resonance
motion.
More accurate estimate could be done with proper simulations of such
effects, but
the total contribution should be smaller than 
 $\la \cos(\phi_a + \phi_b -2\phi_c) \ra \le 10^{-3} \; v_{2,res}
\,v_{2_c}$, where the factor $10^{-3}$ is coming from the estimates of
non-flow azimuthal correlations~\cite{starv1v4}.
Taking all together, one finds the systematic uncertainty in
measurements of $a_\pi$ parameter below one percent level.

In summary, we propose to use mixed harmonics (three particle)
azimuthal correlations to detect strong symmetry violation effects in heavy
ion collisions. Our estimates of the systematic uncertainties
indicate that the measurements of the asymmetry in pion production
with respect to the system angular momentum can be performed at a level better
than one percent (about the lower limit of the effect as predicted by
theory), and probably much better with an additional study and simulations 
of the effects contributing to the systematic uncertainty.

\acknowledgments{
Enlightening discussions with D.~Kharzeev are gratefully acknowledged. 
This work was supported in part by the
U.S. Department of Energy Grant No. DE-FG02-92ER40713.
}

\bibliographystyle{unsrt}
 \end{document}